\documentclass[a4paper]{jpconf}
\usepackage{graphicx}

\usepackage{amstext, amsmath, amssymb}
\usepackage{tikz}
\usetikzlibrary{trees, decorations.pathmorphing, decorations.markings, arrows, shapes, automata, backgrounds, petri, calc, fadings, positioning, patterns}
\usetikzlibrary{external}
\tikzset{external/force remake}\tikzset{external/system call={pdflatex \tikzexternalcheckshellescape -halt-on-error -interaction=batchmode -jobname "\image" "\texsource" && pdftops -eps -f 1 -l 1 "\image".pdf}}
\begin{document}
\title{Dyson\,-\,Schwinger equations in scalar electrodynamics}

\author{A V Berezin$^1$, A A Mironov$^{1,2}$, E S Sozinov$^1$ and A M Fedotov$^1$}

\address{$^1$National Research Nuclear University MEPhI (Moscow Engineering Physics Institute), Moscow, Russia}
\address{$^2$Prokhorov General Physics Institute of the Russian Academy of Sciences, Moscow, Russia} 

\ead{ArsenBRS@mail.ru}

\begin{abstract}
Counterparts of the Dyson\,-\,Schwinger equations for scalar QED in an external electromagnetic field are derived. Exact structure and diagrammatic interpretation of the corresponding mass and polarization operators are obtained. It is shown that in the presence of an external field the final equations include an exact three\,-\,photon interaction vertex.
\end{abstract}

\section{Introduction}
\begin{center}
\begin{figure}[t!]
\centerline{\includegraphics[scale = 1.2]{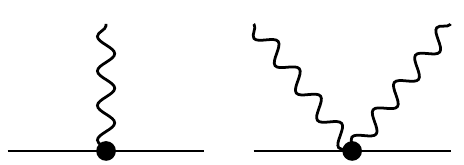}}
\caption{Two types of vertices in scalar QED (the second vertex corresponding to a contact two\,-\,photon interaction is absent in standard fermionic QED).}
\label{pik1}
\end{figure}
\end{center}

Quantum electrodynamics (QED) is a theory of the interacting fermionic electron\,-\,positron and electromagnetic fields. However,  it sometimes might be expedient to isolate or neglect contribution of the spin effects. In addition, there also do exist charged scalar particles. In such situations a scalar version of QED can be used. At first glance, such a theory should be simpler due to the absence of spin degrees of freedom. However, an additional, with respect to the standard fermionic QED, interaction (bare vertex), emerges in such a theory due to the gauge invariance, see Figure~\ref{pik1}.

One of the most important QED equations are the Dyson\,-\,Schwinger (DS) equations \cite{dyson1949s, schwinger1951green}, which establish relationships between the exact (dressed by radiative corrections) propagators and vertices. In particular, it is convenient to use them to construct nonperturbative methods based on partial resummations of the perturbative series.  

In this paper, we derive and discuss the analogs of the DS equations for scalar QED in an external electromagnetic field. They have been already discussed in \cite{binosi2007pinch}, but without a detailed derivation, by only selecting the one\,-\,particle\,-\,irreducible diagrams and assuming that no external field is present. Here we follow a technique described in detail for the standard fermionic QED in \cite{williams2007schwinger}, based on functional integral, which is especially convenient for deriving and analyzing general properties and relations. It is shown that, in the presence of an external field, the DS equations contain an additional contribution with a three\,-\,photon vertex, which was not even mentioned in \cite{binosi2007pinch}.

\section{Notations, basic and auxiliary relations}
We use the Minkowski metric 
$\eta_{\mu\nu} = \text{diag}(+,-,-,-)$, a convention of summation over the repeated indices
$x_\mu y_\mu \equiv x_0 y_0 - \mathbf{xy}$, $x^2 \equiv x_\mu x_\mu$, and a condensed notation for a multifold integral
\begin{equation}
\int d^4z_i f(z_1, \dots ,z_n) \equiv
\int d^4 z_1 \cdots \int d^4z_n f(z_1, \dots ,z_n).
\end{equation}
Unfortunately, there are still no generally accepted unique notations in the literature on quantum field theory. We use the notations consistent with \cite{peskin2018introduction}, whenever possible. 

Consider a generating functional of scalar QED
\begin{equation} \label{2b}
Z[J, \eta, \eta^\dag] = \int \mathcal{D} A \mathcal{D} \varphi^\dag \mathcal{D} \varphi \,
\exp \left( iS[A, \varphi^\dag, \varphi] + i\int d^4y 
\left(  J_\mu A_\mu + \eta^\dag \varphi + \varphi^\dag \eta \right) \right)
\end{equation}
with action
\begin{equation} \label{3n}
S[A, \varphi^\dag, \varphi] = \int d^4y \left(
(D_\mu \varphi)^\dag(D_\mu \varphi) - m^2 \varphi^\dag \varphi - 
\frac{1}{4} F_{\mu\nu}F_{\mu\nu} - \frac{1}{2\xi} \left(\partial_\mu A_\mu\right)^2 \right),
\end{equation} 
where $A_\mu$ is electromagnetic 4\,-\,potential; $\varphi, \varphi^\dag$ are complex scalar field of mass $m$ and its conjugate; $J_\mu$, $\eta^\dag$, and $\eta$ are classical sources of these fields. The last term in \eqref{3n} fixes the gauge, which depends on the value of the parameter $\xi$. The covariant derivative of the charged scalar field and the electromagnetic field tensor are defined as
\begin{equation}
D_\mu = \partial_\mu + ieA_\mu, \qquad
F_{\mu\nu} = \partial_\mu A_\nu - \partial_\nu A_\mu.
\end{equation}
The functional \eqref{2b} has a meaning of the vacuum\,-\,vacuum transition amplitude in the presence of classical sources and fully characterizes the theory. We consider the sources $\eta$, $\eta^\dag$ auxiliary, setting them zero after taking all the required variational derivatives with respect to them, while keeping the current $J_\mu$ arbitrary as generating a (not necessarily self-propagating) external field.    

Strictly speaking, since the one\,-\,loop corrections to the four\,-\,point scalar correlation function diverge logarithmically, for the renormalizability of the theory, the action \eqref{3n} should also include self\,-\,interaction $\lambda (\varphi^\dag \varphi)^2$. The corresponding contributions to the DS equations are described in detail in \cite{binosi2007pinch}. Here, for brevity, we restrict ourselves to a discussion of the account for only the interaction of scalar particles with the electromagnetic field.

The functional
\begin{equation} \label{5b}
W[J, \eta, \eta^\dag] = -i\ln \left( Z[J, \eta, \eta^\dag] \right), 
\end{equation}
generates the connected Green's functions, in particular, one\,-\,point (mean fields)
\begin{equation}
\frac{\delta W}{\delta J_\mu(x)} = A_\mu(x), \quad \label{13}
\frac{\delta W}{\delta \eta^\dag(x)} = \varphi(x), \quad
\frac{\delta W}{\delta \eta(x)} = \varphi^\dag(x),    
\end{equation}
and two\,-\,point (propagators)
\begin{equation} \label{7n}
G(x, y) \equiv -i\left. \frac{\delta^2 W}{\delta \eta^\dag(x) \delta \eta(y)} 
\right|_{\eta^\dag = \eta = 0} \, , \qquad
\mathrm{D}_{\mu\nu}(x,y) \equiv
\left. -i\frac{\delta^2 W}{\delta J_\mu(x) \delta J_\nu(y)} 
\right|_{\eta^\dag = \eta = 0} \, .
\end{equation}
They are functionals of the current $J_\mu$ to account for the presence of an external electromagnetic field. In its absence, in the zeroth order of perturbation theory and in the momentum representation, the propagators \eqref{7n} are reduced to (cf. \cite{peskin2018introduction})
\begin{equation}
G^{(0)}(p) = \frac{i}{p^2-m^2+i0}, \qquad 
\mathrm{D}^{(0)}_{\mu\nu}(k) = - \frac{i}{k^2 +i0} \left( \eta_{\mu\nu} - (1 - \xi)\frac{k_\mu k_\nu}{k^2} \right).
\end{equation}

In virtue of \eqref{5b}, \eqref{13} the following useful representations are also valid for the Green's functions \eqref{7n}:
\begin{equation} \label{10n}
G(x, y) = - \frac{1}{Z} 
\left. \frac{\delta^2 Z}{\delta \eta^\dag(x) \delta \eta(y)} 
\right|_{\eta = \eta^\dag = 0}, \qquad
\mathrm{D}_{\mu\nu}(x,y) = -i \frac{\delta A_\nu(y)}{\delta J_\mu(x)} \,.
\end{equation}

Next, we define an effective action functional by means of the Legendre transform
\begin{equation} \label{6n}
\Gamma[A, \varphi^\dag, \varphi] = W[J, \eta, \eta^\dag] -
\int d^4x \left( J_\mu A_\mu + \eta^\dag \varphi + \varphi^\dag \eta \right).
\end{equation}
This implies the equations for the one\,-\,point functions \eqref{13}:
\begin{equation}
\frac{\delta \Gamma}{\delta A_\mu(x)} = -J_\mu(x), \quad \label{9n}
\frac{\delta \Gamma}{\delta \varphi(x)} = -\eta^\dag(x), \quad
\frac{\delta \Gamma}{\delta \varphi^\dag(x)} = -\eta(x).
\end{equation}
In addition, the second variational derivatives of the functionals \eqref{5b} and \eqref{6n} are related through
\begin{multline} \label{11b}
\int d^4z \left. \frac{\delta^2 W}
{\delta \eta^\dag(x) \delta \eta(z)} \right|_{\eta = \eta^\dag = 0} \times
\left. \frac{\delta^2 \Gamma}{\delta \varphi^\dag(z) \delta \varphi(y)}
\right|_{\varphi^\dag = \varphi = 0} = 
\\
= -\int d^4z \left( \left. \frac{\delta \varphi^\dag(z)}{\delta \eta^\dag(x)} \times
\frac{\delta \eta^\dag(y)}{\delta \varphi^\dag(z)} \right) 
\right|_{\eta = \eta^\dag = 0} = -\delta^4(x - y).
\end{multline} 
Note that resetting the sources in the second line of \eqref{11b} automatically entails vanishing the fields, $\varphi^\dag = \varphi = 0$. According to \eqref{11b}, the kernel of an operator, inverse to the scalar propagator \eqref{7n} reads
\begin{equation} \label{12n}
G^{-1}(x,y) = 
\left. -i\frac{\delta^2 \Gamma}{\delta \varphi^\dag(x) \delta \varphi(y)} 
\right|_{\varphi^\dag = \varphi = 0} \, .
\end{equation}
Similarly, for the kernel of an operator inverse to the photon propagator, we find
\begin{equation} \label{14n}
\mathrm{D}^{-1}_{\mu\nu}(x,y)  = 
\left. -i\frac{\delta^2 \Gamma}{\delta A_\mu(x) \delta A_\nu(y)} 
\right|_{\varphi^\dag = \varphi = 0} \,.
\end{equation}

\section{General approach: a quantum equation of motion}
We recall a general derivation of a quantum equation of motion, an exact functional equation for the generating functional \cite{williams2007schwinger,peskin2018introduction, rivers1988path}. For simplicity, consider a single scalar field $\phi$ and the corresponding source $J$. 
We proceed by noting that a functional integral of variational derivative equals zero identically,
\begin{equation} \label{1}
\int\negthinspace \mathcal{D} \phi \, \frac{\delta F}{\delta \phi(x)} = 0.
\end{equation}

In particular, by setting $F[J, \phi] = \exp \left( iS[\phi] + i\int\negthinspace dx J(x) \phi(x) \right)$, we arrive at the equation
\begin{equation} \label{3}
\int\negthinspace \mathcal{D} \phi
\left( \frac{\delta S}{\delta \phi}(\phi(x)) + J(x) \right) F[J,\phi] = 0.
\end{equation}

To reformulate \eqref{3} in terms of the generating functional
$Z[J] \equiv \int\negthinspace \mathcal{D} \phi \, F[J, \phi]$, we note that for an arbitrary polynomial function $f$ we have
\begin{equation}
\int\negthinspace \mathcal{D} \phi \, f(\phi(x)) F[J,\phi] = 
f \left( -i \frac{\delta}{\delta J(x)} \right) Z[J].
\end{equation}
Hence \eqref{3} can be equivalently represented as
\begin{equation} \label{5}
\left( \frac{\delta S}{\delta \phi}\left( -i \frac{\delta}{\delta J(x)} \right) + J(x) \right) Z[J] = 0.
\end{equation}

The resulting equation \eqref{5} is called the quantum equation of motion, it serves a starting point for deriving the DS equations \cite{williams2007schwinger,peskin2018introduction, rivers1988path}.

\section{Equation for a scalar particle propagator}
In scalar QED, by choosing $\varphi^\dag$ as a variable field, equation \eqref{5} takes the form
\begin{equation} \label{35}
\left( \frac{\delta S}{\delta \varphi^\dag} \left( -i \frac{\delta}{\delta J_\mu(x)}, \, -i \frac{\delta}{\delta \eta(x)}, \,
-i \frac{\delta}{\delta \eta^\dag(x)} \right) + \eta(x) \right) Z = 0,
\end{equation}
where, according to \eqref{3n}, the variational derivative of the action reads
\begin{equation} \label{20n}
\frac{\delta S}{\delta \varphi^\dag} =
\left( -\partial^2 - 2ieA_\mu\partial_\mu  + e^2A^2 - m^2 \right) \varphi.
\end{equation}

In this section, let us use the Landau gauge $\xi = 0$, so that
\begin{equation} \label{8b}
\partial_\mu A_\mu = 0, \qquad 
\partial_\mu \mathrm{D}_{\mu\nu} = 0,
\end{equation}
and the electromagnetic field can be freely taken in and out of the 4\,-\,divergence sign. 

After substituting \eqref{20n}, equation \eqref{35} takes the form
\begin{equation}
\left( \partial^2 + m^2 + 2e\partial_\mu \frac{\delta}{\delta J_\mu(x)} 
+ e^2\frac{\delta^2}{\delta J_\mu(x) \delta J_\mu(x)} \right) 
\frac{\delta Z}{\delta \eta^\dag(x)} - 
i\eta(x) Z = 0.
\end{equation}
To obtain an equation for the Green's function \eqref{10n}, we take an additional variational derivative with respect to $\eta (x')$:
\begin{multline} \label{1y}
\left( \partial^2 + m^2 + 2e\partial_\mu \frac{\delta}{\delta J_\mu(x)} 
+ e^2\frac{\delta^2}{\delta J_\mu(x) \delta J_\mu(x)} \right) 
\frac{\delta^2 Z}{\delta \eta^\dag(x) \delta \eta(x')} = 
\\
= i \eta(x) \frac{\delta Z}{\delta \eta^\dag(x')} + i \delta^4(x - x') Z.
\end{multline} 

Taking into account relations \eqref{13}, we transform the variational derivative with respect to $J_\mu$ as follows: 
\begin{equation} \label{42}
\frac{\delta}{\delta J_\mu(x)} = \frac{\delta}{\delta J_\mu(x)} \,
\frac{e^{iW}}{Z} = e^{iW} \left(
i\frac{\delta W}{\delta J_\mu(x)} + \frac{\delta}{\delta J_\mu(x)} \right)
\frac{1}{Z} =
Z \left( iA_\mu (x) + 
\frac{\delta}{\delta J_\mu(x)} \right) \frac{1}{Z} \, .
\end{equation}

\begin{center}
\begin{figure}[t!]
\centerline{\includegraphics[scale = 1]{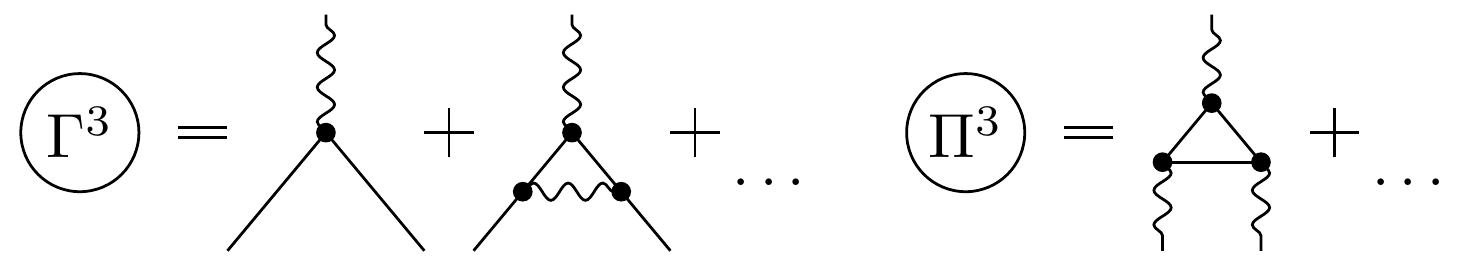}}\vspace{0.2cm}
\centerline{\includegraphics[scale = 1]{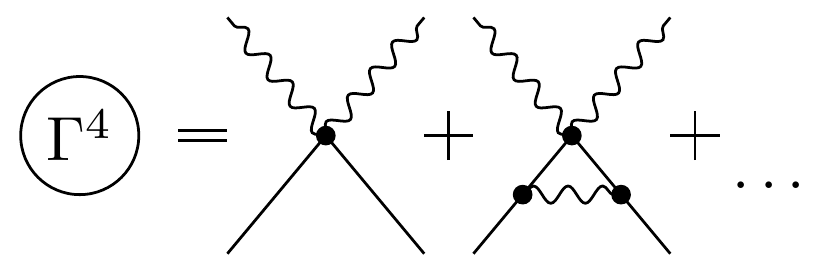}}
\caption{Lowest\,-\,order contributions to the exact vertices $\Gamma^3$, $\Pi^3$ and $\Gamma^4$ (the outer lines are shown for clarity and are not included into the definition of vertices).}
\label{pik4}
\end{figure}
\end{center}

Second derivative is transformed the same way, using representation \eqref{10n} for the photon propagator:
\begin{equation} \label{43}
\frac{\delta^2}{\delta J_\mu(x) \delta J_\mu(x)} = 
Z \left( \frac{\delta^2}{\delta J_\mu(x) \delta J_\mu(x)} +
2iA_\mu(x) \frac{\delta}{\delta J_\mu(x)} - A^2(x) - \mathrm{D}_{\mu\mu}(x, x) 
\right) \frac{1}{Z} \, .
\end{equation}

Substituting \eqref{42}, \eqref{43} into \eqref{1y} and setting $\eta = \eta^\dag = 0$, we arrive at the equation for the scalar propagator
\begin{multline} \label{47}
i\left( \partial^2 + m^2 + 2ie\partial_\mu A_\mu(x)  - e^2A^2(x) - 
e^2 \mathrm{D}_{\mu\mu}(x, x) \right. 
\\
+ \left. 2eD_\mu \frac{\delta}{\delta J_\mu(x)} + 
e^2 \frac{\delta^2}{\delta J_\mu(x) \delta J_\mu(x)}\right)
G(x, x') = \delta^4(x - x').
\end{multline}
It remains to calculate the variational derivatives with respect to the current $J_\mu$ that are contained in \eqref{47}. To do it, we first pass from them to the derivative with respect to the field $A_\nu$:
\begin{equation} \label{26b}
\frac{\delta G(x,x')}{\delta J_\mu(x)} =
\int d^4z_1 
\frac{\delta A_\nu(z_1)}{\delta J_\mu(x)} \times \frac{\delta G(x, x')}{\delta A_\nu (z_1)}\, ,
\end{equation}
by using the total derivative formula and the fact that in scalar QED all odd order variational derivatives of the generating functional with respect to scalar sources $\eta$ and $\eta^\dag$ vanish at $\eta = \eta^\dag = 0$. Furthermore, in virtue of the matrix identity $\delta G = -G \left( \delta G^{-1} \right) G$ and representations \eqref{12n}, \eqref{10n} for the kernels of the inverse scalar and photon propagators, we obtain:
\begin{equation} \label{48}
\frac{\delta G(x,x')}{\delta J_\mu(x)} = i\int d^{4} z_i \mathrm{D}_{\mu\nu}(x, z_1)
G(x, z_2) \Gamma^3_\nu(z_1, z_2, z_3) G(z_3, x'),
\end{equation}
where the  exact three\,-\,tailed vertex (see Figure~\ref{pik4}) is defined by
\begin{equation} \label{28n}
\Gamma^3_\alpha(x, y, z) \equiv i
\left. \frac{\delta^3 \Gamma} {\delta A_\alpha (x) \delta \varphi^\dag(y) \delta \varphi(z)} 
\right|_{\varphi^\dag = \varphi = 0} \, .
\end{equation}

The variational derivative of the photon propagator with respect to $J_\mu$ is calculated in a similar way:
\begin{equation} \label{29b}
\frac{\delta \mathrm{D}_{\mu\nu}(x,x')}{\delta J_\mu(x)} = i\int d^{4} z_i \mathrm{D}_{\mu\alpha}(x, z_1)
\mathrm{D}_{\mu\beta}(x, z_2) \Pi^3_{\alpha\beta\gamma}(z_1, z_2, z_3) \mathrm{D}_{\gamma\nu} (z_3, x'),
\end{equation}
where the exact three\,-\,photon vertex (see Figure~\ref{pik4}) is defined by
\begin{equation} \label{30b}
\Pi^3_{\alpha\beta\gamma}(x, y, z) \equiv i
\left. \frac{\delta^3 \Gamma} {\delta A_\alpha (x) \delta A_\beta (y) \delta A_\gamma (z)} 
\right|_{\varphi^\dag = \varphi = 0} \, .
\end{equation}

\begin{center}
\begin{figure}[t!]
\centerline{\includegraphics[scale = 1.3]{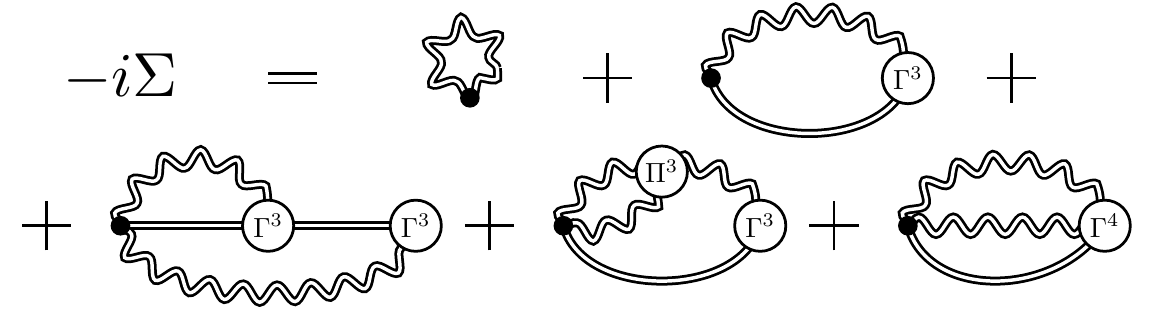}}
\caption{Diagrammatic representation for the mass operator \eqref{34n}.}
\label{pik2}
\end{figure}
\end{center}

To find the second variational derivative of the scalar propagator in \eqref{47} we take a derivative of \eqref{48} using relations \eqref{48} - \eqref{29b}. The result contains a four\,-\,tailed vertex (see Figure~\ref{pik4})
\begin{equation} \label{31n}
\Gamma^4_{\alpha\beta}(x, y, z, w) \equiv i
\left. \frac{\delta^4 \Gamma}
{\delta A_\alpha(x) \delta A_\beta(y) \delta \varphi^\dag(z) \delta \varphi(w)} 
\right|_{\varphi^\dag = \varphi = 0} \, .
\end{equation}

After that, equation \eqref{47} is finally cast to the form
\begin{equation} \label{32b}
(D^2 +  m^2)G(x,x') + \int d^4y \Sigma(x, y) G(y,x') = -i\delta^4(x-x'),
\end{equation}
similar to the DS equation in standard QED. The mass operator having appeared in \eqref{32b} reads (see Figure~\ref{pik2})
\begin{align}
-i\Sigma(x, y) = & \, ie^2 \mathrm{D}_{\mu\mu}(x, x)\delta^4(x-y) +
2e \int d^{4} z_i \mathrm{D}_{\mu\nu}(x, z_1)
\left( D_\mu G(x, z_2) \right) \Gamma^3_\nu(z_1, z_2, y)  
\notag
\\
& + 2ie^2\int d^{4} z_i \mathrm{D}_{\mu\nu}(x, z_1) \left( 
\int d^{4} w_i \mathrm{D}_{\mu\nu}(x, w_1)
G(x, w_2) \Gamma^3_\nu(w_1, w_2, w_3) G(w_3, z_2)
\right) 
\notag
\\
& \mspace{20mu} \times \Gamma^3_\nu(z_1, z_2, y) 
\notag
\\
& + ie^2\int d^{4} z_i \left( 
\int d^{4} w_i \mathrm{D}_{\mu\alpha}(x, w_1)
\mathrm{D}_{\mu\beta}(x, w_2) \Pi^3_{\alpha\beta\gamma}(w_1, w_2, w_3) 
\mathrm{D}_{\gamma\nu}(w_3, z_1) \right)
\notag
\\
& \mspace{20mu} \times G(x, z_2) \Gamma^3_\nu(z_1, z_2, y) 
\notag
\\ 
& + \left. ie^2\int d^{4} z_i \mathrm{D}_{\mu\nu}(x, z_1)
G(x, z_2) \left(  \int d^4w \mathrm{D}_{\mu\alpha}(x, w) 
\Gamma^4_{\alpha\nu}(w, z_1, z_2, y) \right) \right). \label{34n}
\end{align}

Expression \eqref{34n} coincides up to notation with the result discussed in \cite{binosi2007pinch}, except for that \eqref{34n} does not contain the vertices induced by the self\,-\,interaction of the scalar field, but contains the additional contribution of the three\,-\,photon vertex $\Pi^3_{\alpha \beta \gamma}$. In the absence of an external field, it vanishes identically as any C-odd operator in a C-invariant theory (a scalar analog of Furry's theorem). However, in the presence of an external field, the C-invariance is violated, as a result the three\,-\,photon vertex not only remains, but even generates on\,-\,shell processes of photon splitting and merging \cite{adler1971photon}. In other words, since the exact scalar propagators contain contributions of an arbitrary number of interactions with the external field, the total number of vertices in a scalar loop with an odd number of outer photon lines is never actually limited in this case to an odd value.

We also note that the incoming scalar outer lines are expressed in terms of the two\,-\,point functions $G(x,x')$ for $x_0'\to-\infty$, therefore satisfy the homogeneous version of DS equation \eqref{32b} with zero right\,-\,hand side.

\section{Equation for a photon propagator}
Let us proceed to obtaining an equation for the exact photon propagator. We start again with the identity \eqref{5}, where now we take $A_\mu$ as a variable field:
\begin{equation} \label{55}
\left( \frac{\delta S}{\delta A_\mu} \left( -i \frac{\delta}{\delta J_\mu(x)}, \, -i \frac{\delta}{\delta \eta(x)}, \,
-i \frac{\delta}{\delta \eta^\dag(x)} \right) + J_\mu(x) \right) Z = 0.
\end{equation}
Taking into account \eqref{3n}, we have
\begin{equation}
\frac{\delta S}{\delta A_\mu} =
-ie \left( \varphi^\dag \partial_\mu \varphi - \varphi \partial_\mu \varphi^\dag \right) +
2e^2A_\mu \varphi^\dag \varphi +
\partial^2 A_\mu - (1 - \xi^{-1}) \partial_\mu \partial_\nu A_\nu.
\end{equation}
Therefore, equation \eqref{55} takes the form:
\begin{multline} \label{61}
\left( -ie \lim_{x'\rightarrow x}(\partial'_\mu - \partial_\mu) \frac{\delta^2}{\delta \eta^\dag(x) \delta \eta(x')} + 2ie^2 \frac{\delta^3}{\delta J_\mu(x) \delta \eta^\dag(x) \delta \eta(x)} 
\right.
\\
- \left. i \partial^2 \frac{\delta}{\delta J_\mu(x)} + 
i(1 - \xi^{-1}) \partial_\mu \partial_\nu 
\frac{\delta}{\delta J_\nu(x)} + J_\mu(x) \right) Z = 0,
\end{multline}
where, for the convenience of further calculations, in the first term the variational derivatives with respect to currents $\eta$, $\eta^\dag$ are regularized by point splitting. 

As in derivation of the equation for the scalar propagator, one could then take an extra variational derivative with respect to $J_\nu$. However, in this case it is easier to derive an equation for the inverse propagator. To this end, we first transform \eqref{61} using \eqref{13} and rewriting the higher\,-\,order variational derivatives of the generating functional as follows,
\begin{equation} \label{63}
\frac{\delta^2 Z}{\delta \eta^\dag(x) \delta \eta(x')} = e^{iW} \left(
- \varphi(x) \varphi^\dag(x') + 
i\frac{\delta^2 W}{\delta \eta^\dag(x) \delta \eta(x')} \right),
\end{equation} 

\begin{multline} \label{64}
\frac{\delta^3 Z}{\delta J_\mu(x) \delta \eta^\dag(x) \delta \eta(x)} = e^{iW} \left( -iA_\mu(x) \varphi(x) \varphi^\dag(x) -
A_\mu(x) \frac{\delta^2 W}{\delta \eta^\dag(x) \delta \eta(x)} \right.
\\
- \left. \varphi^\dag(x) \frac{\delta^2 W}{\delta J_\mu(x) \delta \eta^\dag(x)} -
\varphi(x) \frac{\delta^2 W}{\delta J_\mu(x) \delta \eta(x)} +
i\frac{\delta^3 W}{\delta J_\mu(x)\delta \eta^\dag(x) \delta \eta(x)} \right).
\end{multline}
Next, we set in \eqref{61} - \eqref{64} $\eta = \eta^\dag = 0$, which implies $\varphi^\dag = \varphi = 0$ and therefore nullifies most of the terms on the right\,-\,hand sides of the formulas \eqref{63}, \eqref{64}, while the second derivatives of $W$ with respect to scalar sources are expressed through the two\,-\,point function Green according to \eqref{7n}. After substituting them into \eqref{61} and expressing the current in terms of the effective action according to \eqref{9n}, we arrive at the equation
\begin{multline} \label{40b}
\left. \frac{\delta \Gamma}{\delta A_\mu(x)} \right|_{\varphi^\dag = \varphi = 0} = 
ie \lim_{x'\rightarrow x}(\partial'_\mu - \partial_\mu) G(x,x') +
2e^2 A_\mu(x) G(x,x) - 2ie^2 \frac{\delta}{\delta J_\mu(x)} G(x,x) 
\\
+ (\eta_{\mu\alpha} \partial^2  - (1 - \xi^{-1}) \partial_\mu \partial_\alpha ) A_\alpha(x) = -J_\mu(x).
\end{multline}
The resulting equation \eqref{40b} has the meaning of an equation for the one\,-\,point Green's function of the electromagnetic field (mean field) and can serve for deriving an effective electromagnetic action (a generalization of the Heisenberg\,-\,Euler one\,-\,loop effective action with all the multi\,-\,loop corrections taken into account).

To find the kernel of the inverse propagator \eqref{14n} it suffices to differentiate \eqref{40b} with respect to $A_\nu(y)$:
\begin{multline} \label{42n}
\mathrm{D}_{\mu\nu}^{-1}(x,y) = 
-i(\eta_{\mu\nu}\partial^2  - (1 - \xi^{-1}) \partial_\mu \partial_\nu ) \delta^4(x-y) -
2ie^2 \eta_{\mu\nu} \delta^4(x - y) G(x,x) 
\\
- e \lim_{x'\rightarrow x} \left( D_\mu- (D_\mu')^\dag \right) 
\frac{\delta G(x,x')}{\delta A_\nu(y)} - 
2e^2 \frac{\delta^2 G(x,x)}{\delta J_\mu(x) \delta A_\nu(y)} \,.
\end{multline}
Next, we express the variational derivative with respect to $A_\nu$ in the same way as in \eqref{26b}:
\begin{equation} \label{43n}
\frac{\delta G(x,x')}{\delta A_\nu(y)} =
\int d^4 z_i \, G(x,z_1) \Gamma^3_\nu(y,z_1,z_2) G(z_2,x').
\end{equation}

\begin{center}
\begin{figure}[t!]
\centerline{\includegraphics[scale = 1.3]{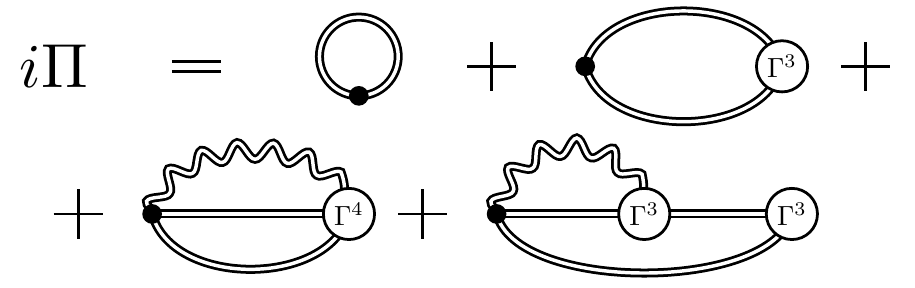}}
\caption{Diagrammatic representation for the polarization operator \eqref{pi_f}.}
\label{pik3}
\end{figure}
\end{center}

Substituting \eqref{43n} into \eqref{42n} and transforming the remaining variational derivative with respect to the current $J_\mu$ by using \eqref{48}, \eqref{28n}, \eqref{31n}, we obtain
\begin{align}\label{d-1}
\mathrm{D}_{\mu\nu}^{-1}(x,y) = -i(\eta_{\mu\nu}\partial^2  - (1 - \xi^{-1}) \partial_\mu \partial_\nu ) \delta^4(x-y) - i\Pi_{\mu\nu}(x,y),
\end{align}
with the polarization operator $\Pi_{\mu\nu}(x,y)$ given by the expression (see Figure~\ref{pik3})
\begin{align}\label{pi_f}
i\Pi_{\mu\nu}(x,y) = & \, 2ie^2 \eta_{\mu\nu} \delta^4(x - y) G(x,x)
\notag
\\
& + e \lim_{x' \rightarrow x} \left( D_\mu - (D_\mu')^\dag \right) \int d^4 z_i \, 
G(x,z_1) \Gamma^3_\nu(y,z_1,z_2) G(z_2,x') 
\notag
\\
& + 2ie^2 \int d^4 z_i \, \mathrm{D}_{\mu\alpha}(x,z_3) G(x,z_1)
\Gamma^4_{\alpha\nu}(z_3,y,z_1,z_2) G(z_2,x) 
\notag
\\
& + 2ie^2 \int d^4 z_i \left( \int d^{4} w_i \mathrm{D}_{\mu\nu}(x, w_1)
G(x, w_2) \Gamma^3_\nu(w_1, w_2, w_3) G(w_3, z_1) \right) 
\notag
\\
& \mspace{20mu} \times \Gamma^3_{\nu}(y, z_1, z_2)
G(z_2,x).
\end{align}
The polarization operator \eqref{pi_f} is transverse $\partial_\mu\Pi_{\mu\nu}=0$, as follows from the Ward identity and can be verified directly. By taking a 4\,-\,divergence of \eqref{d-1}, with account for definition of $\mathrm{D}^{-1}$ as an inverse operator, we further derive
\begin{align}\label{dD}
\partial_\nu\mathrm{D}_{\nu\rho}(x,y)=i\xi\partial^{-2}\partial_\rho\delta^4(x - y).
\end{align}

Finally, by multiplying \eqref{d-1} by $\mathrm{D}_{\nu\rho}(y,x'')$, integrating over the variable $y$, excluding the 4\,-\,divergence of the propagator using \eqref{dD} and turning $\xi\to 0$ at the end, we arrive at an analogue of the DS equation for the photon propagator in the Landau gauge:
\begin{equation}\label{sds}
\partial^2 \mathrm{D}_{\mu \rho}(x,x'') +
\int d^4y \Pi_{\mu\nu}(x,y) \mathrm{D}_{\nu\rho}(y,x'') = 
i (\eta_{\mu \rho} - \partial_\mu \partial_\rho / \partial^2) \delta^4(x-x''),
\end{equation}

As for scalar particles, the external incoming photon line satisfies the homogeneous version of the resulting equation \eqref{sds}. 

\section{Conclusion}
Counterparts of the DS equations in scalar QED in an external electromagnetic field are obtained and a diagrammatic representations for the corresponding mass and polarization operators are constructed. It is shown that they have a much more complex form than in standard QED. In particular, in the presence of an external field, the mass operator contains a contribution with an exact three\,-\,photon vertex, the fact not previously pointed out in the literature. The resulting equations can be used for nonperturbative calculations in scalar QED involving partial resummations of a perturbative series.   

\ack{The authors are grateful to K.A. Kazakov for a discussion of the subtle aspects of scalar QED. This work was supported by the Russian Foundation for Basic Research (projects 19-02-00643, 19-32-60084, and 20-52-12046).}

\section*{References}
\providecommand{\newblock}{}


\begin{thebibliography}{1}
\expandafter\ifx\csname url\endcsname\relax
  \def\url#1{{\tt #1}}\fi
\expandafter\ifx\csname urlprefix\endcsname\relax\def\urlprefix{URL }\fi
\providecommand{\eprint}[2][]{\url{#2}}

\bibitem{dyson1949s}
Dyson F~J 1949 {\em Physical Review\/} {\bf 75} 1736

\bibitem{schwinger1951green}
Schwinger J 1951 {\em Proceedings of the National Academy of Sciences\/} {\bf
  37} 452--455

\bibitem{binosi2007pinch}
Binosi D and Papavassiliou J 2007 {\em Journal of High Energy Physics\/} {\bf
  2007} 041

\bibitem{williams2007schwinger}
Williams R 2007 {\em Schwinger-Dyson equations in QED and QCD the calculation
  of fermion-antifermion condensates\/} Ph.D. thesis Durham University

\bibitem{peskin2018introduction}
Peskin M 2018 {\em An introduction to quantum field theory\/} (CRC press)

\bibitem{rivers1988path}
Rivers R 1988 {\em Path integral methods in quantum field theory\/} (Cambridge
  University Press)

\bibitem{adler1971photon}
Adler S~L 1971 {\em Annals of Physics\/} {\bf 67} 599--647
\end{thebibliography}
\end{document}